\documentclass[%
 reprint, superscriptaddress,
 amsmath,amssymb,
 aps,
]{revtex4-2}

\usepackage{graphicx}
\usepackage{dcolumn}
\usepackage{bm}
\usepackage{siunitx}

\usepackage[normalem]{ulem}

\usepackage{xcolor}
\definecolor{amber}{rgb}{1,0.49,0}

\newcommand{\editor}[2]{%
	\expandafter\newcommand\csname #1note\endcsname[1]{%
		\textcolor{#2}{(\textbf{#1:} ##1)}}%
	\expandafter\newcommand\csname #1\endcsname[1]{%
		\textcolor{#2}{##1}}%
	\expandafter\newcommand\csname #1cancel\endcsname[1]{%
		\textcolor{#2}{\sout{##1}}}%
	\expandafter\newcommand\csname #1change\endcsname[2]{%
		\textcolor{#2}{\sout{##1} ##2}}%
	\expandafter\newcommand\csname #1prov\endcsname[2]{%
		\textcolor{#2}{[(##1) (##2)]}}%
	\newenvironment{#1text}{\color{#2}}{\color{black}}
}

\editor{FP}{blue}
\begin{document}

\preprint{APS/123-QED}

\title{A Fresh Perspective on Water Dynamics in Aqueous Salt Solutions}


\author{Rolf Zeißler}
\affiliation{%
Institute for Condensed Matter Physics, Technical University of Darmstadt, D-64289 Darmstadt, Germany
}%
\author{Florian Pabst}
\affiliation{%
SISSA—Scuola Internazionale Superiore di Studi Avanzati, 34136 Trieste, Italy
}%
\author{Thomas Blochowicz}
\affiliation{%
Institute for Condensed Matter Physics, Technical University of Darmstadt, D-64289 Darmstadt, Germany
}%


\date{\today}

\begin{abstract}
\noindent
Molecular dynamics in pure water and aqueous salt solutions remain incompletely understood, partly due to the apparent contradictions between results from different spectroscopic techniques.
In this work, we demonstrate, by detailed comparison of light scattering and dielectric spectroscopy data for pure water and aqueous lithium chloride solutions, that these apparent contradictions can be resolved by accounting for orientational cross-correlations of neighboring molecules. Remarkably, a single structural relaxation mode with largely temperature- and concentration-independent shape can be identified in all spectra, from room temperature down to the deeply supercooled regime. These results provide a new perspective for the study of molecular dynamics in aqueous salt solutions.
\end{abstract}

\maketitle

\noindent
Understanding molecular dynamics of water and especially aqueous solutions is of utmost importance for unraveling the complex nature of the physics on which all living organisms are based. However, even for rather simple aqueous salt solutions, this dynamics is only partly understood. One factor that has hindered understanding are apparent contradictions between results from different spectroscopic techniques, e.g., broadband dielectric spectroscopy (BDS)\cite{barthel1990dielectric,ronne2002low,buchner1999dielectric,ishai2015primary,kaatze2002hydrogen,popov2016mechanism}, depolarized dynamic light scattering (DDLS)\cite{sokolov1995dynamics, hansen2016identification, fukasawa2005relation,nakanishi2012no}, nuclear magnetic resonance spectroscopy (NMR)\cite{schneider2018nmr} and quasielastic neutron scattering (QENS)\cite{qvist2011structural, arbe2016dielectric}.\\
In this work, we focus on resolving the contradictions between two of these methods, namely BDS and DDLS. Although both techniques monitor molecular motion, the former via the electric dipole moment \cite{kremer2002broadband} and the latter via the polarizability tensor \cite{berne2000dynamic}, a huge difference between BDS and DDLS spectra becomes apparent from studies on pure water. Several groups reported a difference of at least one order of magnitude between the positions of the main relaxation peak of dielectric loss and DDLS susceptibility spectra \cite{fukasawa2005relation,hansen2016identification,popov2016mechanism}. Furthermore, the overall shape of the relaxation spectra differs strongly. Where the dielectric loss spectra are dominated by a Debye-like peak centered around \SI{20}{\giga\hertz}, the DDLS spectra exhibit a relaxation process which is significantly less strong and shows up only as a shoulder on the vibrational part of the spectrum. In some of the aforementioned works, this discrepancy between the spectra obtained by the two methods has been interpreted as arising from a strong Debye-like relaxation process connected to long-range dipolar interactions introduced by the hydrogen bond network, which only contribute to the dielectric loss and not to the DDLS spectra \cite{fukasawa2005relation,hansen2016identification}. The actual structural relaxation contribution is suspected to be hidden below the strong Debye process in BDS.\\
Confirming results regarding the nature of the BDS spectrum recently came from classical molecular dynamics (MD) simulations. Alvarez et al. \cite{alvarez2023debye} showed that dipolar cross-correlations make up for approximately \SI{60}{\percent} of the dielectric response of water and the total correlation function therefore decays on a longer timescale than the self part of the dipolar correlation function. Similar results were obtained by ab initio molecular dynamics simulations \cite{holzl2021dielectric}. Since DDLS is not sensitive to orientational cross correlations in most cases \cite{bohmer2024spectral, pabst2021generic, pabst2020dipole, gabriel2020intermolecular}, this could explain some of the discrepancies between the two methods. In fact, it was recently shown that the dielectric loss of polar liquids can be strongly affected by slow dipolar cross-correlation contributions \cite{bohmer2024dipolar}. For water, however, experimental validation is difficult due to the highly superimposed nature of the different contributions. A possible route has recently opened up, inspired by first-principles simulations on aqueous sodium chloride solutions. Zhang et al. \cite{zhang2023dissolving} reported a decrease of the Kirkwood correlation factor $g_{\mathrm{k}}$, which represents a measure of the prevalence of alignment of dipoles in a liquid, with increasing salt concentration. The results show that this decrease of $g_{\mathrm{k}}$ is mediated by the perturbation of the hydrogen bond network by ionic hydration shells.\\ 
Since the slow dipolar cross-correlation contributions to the dielectric response have been shown to scale with $g_{\mathrm{k}}$, BDS and DDLS spectra should become more similar upon addition of salt. However, in aqueous salt solutions the situation could be even more complex. In a recent work by Gonz{\'a}lez-Jim{\'e}nez et al., optical Kerr-effect (OKE) spectra of aqueous salt solutions were interpreted as consisting of a slow hydration water and a fast bulk-like water mode \cite{gonzalez2023lifting}. For BDS spectra, such a contribution is reported only rarely and is shown to be very weak \cite{buchner2023ion}. The present study aims to elucidate the origin of the BDS and DDLS spectra of aqueous salt solutions, by a detailed comparison of the results from both experimental techniques, gathered over a broad temperature and frequency range for pure water and aqueous lithium chloride solutions.\\
The aqueous lithium chloride solution at \SI{14.8}{\mol\percent} ($8\,\mathrm{M}$) salt concentration was purchased from Sigma Aldrich. This concentration lies within the eutectic range of solutions of lithium chloride and water but is still well below the solubility limit \cite{prevel1995structural}. All other concentrations were produced by diluting this stock solution with deionized MilliQ water, which was also used for the measurements on pure water. Depolarized dynamic light scattering (DDLS) spectra were obtained by three different experimental setups with different frequency ranges, i.e., photon correlation spectroscopy (PCS) ($\sim$\SI{}{\milli\hertz}-\SI{}{\mega\hertz}),
\begin{figure}[ht!]
  \centering
  \includegraphics[width=0.5\textwidth]{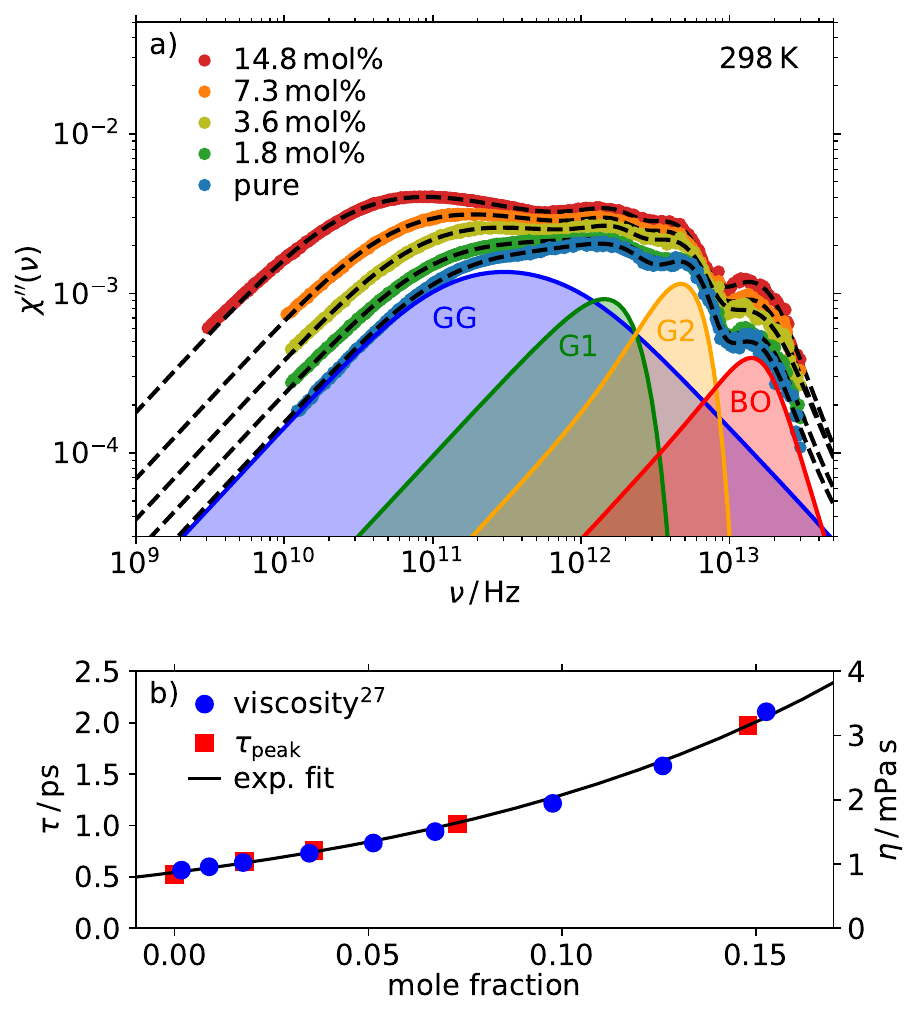}
  \caption{a) DDLS (combined TFPI and RS) spectra of pure water and aqueous lithium chloride solutions at \SI{298}{\kelvin}. Colored curves are model functions superimposed to describe the data (see text). The black dashed curves represent the total fit function. b) Peak relaxation times of the relaxation peak compared to literature data of the viscosity at \SI{298}{\kelvin} by Tanaka et al. \cite{tanaka1991physico}.}
  \label{fig:figure1}
\end{figure}
tandem Fabry-Perot interferometry (TFPI) and Raman spectroscopy (RS) ($\sim$\SI{}{\giga\hertz}-\SI{}{\tera\hertz}). The experimental setups are described in detail elsewhere \cite{gabriel2018depolarized}. Throughout this work the DDLS data will be presented as the imaginary part of the generalized DDLS susceptibility $\chi^{\prime\prime}(\nu)$. The dielectric data were obtained by two different experimental setups. One employing frequency response measurements, with a Novocontrol Alpha-N analyzer and parallel plate capacitor as a sample cell ($\sim$\SI{}{\milli\hertz}-\SI{}{\mega\hertz}) and another employing microwave reflectometry, with an Agilent PNA-L N5230C network analyzer and slim form probe of the dielectric probe kit ($\sim$\SI{100}{\mega\hertz}-\SI{50}{\giga\hertz}). The dielectric data will be presented in terms of the imaginary part of the dielectric permittivity, i.e., the dielectric loss $\epsilon^{\prime\prime}(\nu)$, throughout this work. For all presented spectra the DC conductivity contribution $\epsilon^{\prime\prime}_{\mathrm{DC}}(\nu)$ has been subtracted, i.e., the presented loss is $\epsilon^{\prime\prime}(\nu)=\epsilon^{\prime\prime}_{\mathrm{total}}(\nu)-\epsilon^{\prime\prime}_{\mathrm{DC}}(\nu)$. This allows for a direct comparison of the data obtained by both methods. Concentration dependent measurements were performed at $\SI{298}{\kelvin}$ and $\SI{263}{\kelvin}$ in both DDLS and BDS. For the solution at \SI{14.8}{\mol\percent} also measurements in the supercooled regime (\SI{145}{\kelvin}$-\SI{160}{\kelvin}$) were performed.\\
Fig.\,\ref{fig:figure1} a) shows DDLS spectra at $\SI{298}{\kelvin}$ of pure water and aqueous lithium chloride solutions at different concentrations. The spectra were obtained in a way that relative amplitudes are accurate and the data are not subjected to any form of scaling \cite{pabst2022intensity}.
\begin{figure}[ht]
  \centering
  \includegraphics[width=0.5\textwidth]{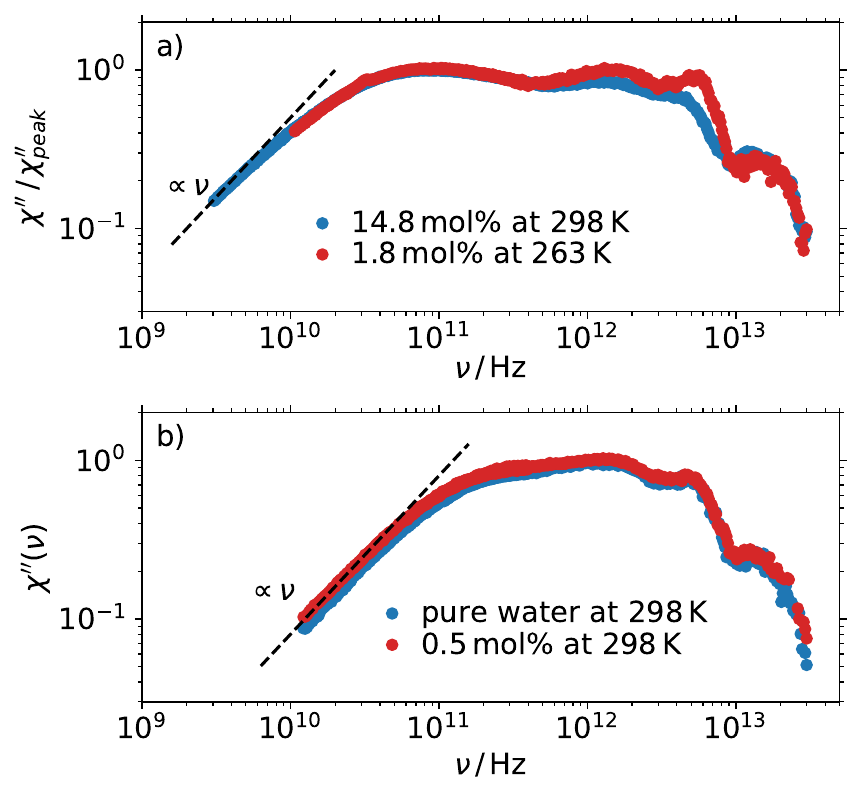}
  \caption{a) DDLS spectra of the \SI{14.8}{\mol\percent} solution at \SI{298}{\kelvin} and the \SI{1.8}{\mol\percent} at \SI{263}{\kelvin} normalized to the amplitude of the relaxation peak. b) DDLS spectra of pure water and the \SI{0.5}{\mol\percent} solution at \SI{298}{\kelvin}.}
  \label{fig:figure2}
\end{figure}
The spectra contain all previously reported features, i.e., the
vibrational contributions above \SI{1}{\tera\hertz} and the relaxation peak ($\nu_{\mathrm{peak}}\approx\SI{200}{\giga\hertz}$ for pure water). As reported in \cite{gonzalez2023lifting}, for OKE spectra, the relaxation peak becomes more pronounced with increasing salt concentration. To describe the data we employed a fit model, which is similar to models that were previously used in the literature to describe DDLS spectra in the \SI{}{\giga\hertz} to \SI{}{\tera\hertz} range. Analogous to ref.\,\cite{gonzalez2023lifting}, a damped harmonic oscillator was used to describe the contribution at $\sim\SI{17}{\tera\hertz}$ and two antisymmetrized gaussian peak functions were used to account for the contributions at $\sim\SI{1.5}{\tera\hertz}$ and $\sim\SI{5}{\tera\hertz}$. However, in contrast to ref.\,\cite{gonzalez2023lifting}, only a single model function was employed to describe the relaxation peak. Here we chose a generalized gamma distribution (GG) \cite{blochowicz2003susceptibility} with a width parameter of $\alpha=2.0$ and a power law exponent of $\beta=0.5$. This model was chosen since it is known to well describe the relaxation peaks of glass forming liquids in the supercooled regime \cite{pabst2021generic} and it will be demonstrated later in this work that this holds true also for aqueous lithium chloride solutions in the supercooled regime. A more detailed description of the fitting procedure and comparison to a fit model employing a Cole-Davidson function to describe the relaxation peak can be found in the supporting information (SI) of this work.\\
Fig.\,\ref{fig:figure1} b) shows the obtained peak relaxation times in dependence of the salt concentration in comparison to viscosity data from the literature \cite{tanaka1991physico}. The obtained peak relaxation times closely follow the viscosity, i.e., $\tau \propto \eta$. This and the fact, that for none of the presented spectra in Fig.\,\ref{fig:figure1} a) a bimodality of the relaxation peak is observed, leads us to believe that there is no need to introduce distinct contributions for hydration- and bulk-like water, in contrast to what has been suggested in the literature \cite{gonzalez2023lifting, turton2011structure,van2023dynamics}.\\
Fig.\,\ref{fig:figure2} a) shows DDLS spectra of the \SI{14.8}{\mol\percent} solution at \SI{298}{\kelvin} and the \SI{1.8}{\mol\percent} solution at \SI{263}{\kelvin}, normalized to the amplitude of the relaxation peak. Without any scaling with respect to frequency, the position and shape of the relaxation peaks match rather well. This result suggests that the peak position simply shifts with changing viscosity, without much change of the relaxation shape and regardless whether the change in viscosity is caused by temperature variation or variation of salt concentration. Interestingly, the relative amplitudes of the vibrational peaks at $\sim\SI{1.5}{\tera\hertz}$ and $\sim\SI{5}{\tera\hertz}$ decrease with increasing salt concentration. Considering the supposed origin of these peaks as intermolecular vibrations between hydrogen-bonded water molecules \cite{walrafen1966raman}, it comes as no surprise that the relative amplitude of these contributions changes due to the perturbation of the hydrogen bond network.\\
Fig.\,\ref{fig:figure2} b) shows DDLS spectra of pure water and a solution of \SI{0.5}{\mol\percent} ($\sim0.25\,\mathrm{m}$) salt concentration at \SI{298}{\kelvin}. The spectra have not been subjected to any form of scaling. Strikingly, apart from a slight shift of the relaxation peak to lower frequencies and an overall slight increase of amplitude, the spectra are almost identical. This, again, contradicts the interpretation of the relaxation peak in DDLS spectra of aqueous salt solutions as a superposition of bulk water and hydration water processes, which are supposed to be similar in amplitude already at such low concentrations (see SI of ref.\,\cite{gonzalez2023lifting}). The dashed lines in Fig.\,\ref{fig:figure2} a) and b) demonstrate that the spectra perfectly follow a linear frequency dependence for $\omega\tau\ll 1$. This additionally proves that there is no need to account for dispersion on the low frequency side of the relaxation peak by employing, e.g., a Cole-Cole or Havriliak-Negami model. Furthermore, we do not observe any shoulder or additional slow process on the low frequency side of the relaxation peak for pure water and any of the solutions, in contrast to a previous DDLS result obtained for pure water \cite{hansen2016identification}.\\
\begin{figure}[h]
  \centering
  \includegraphics[width=0.5\textwidth]{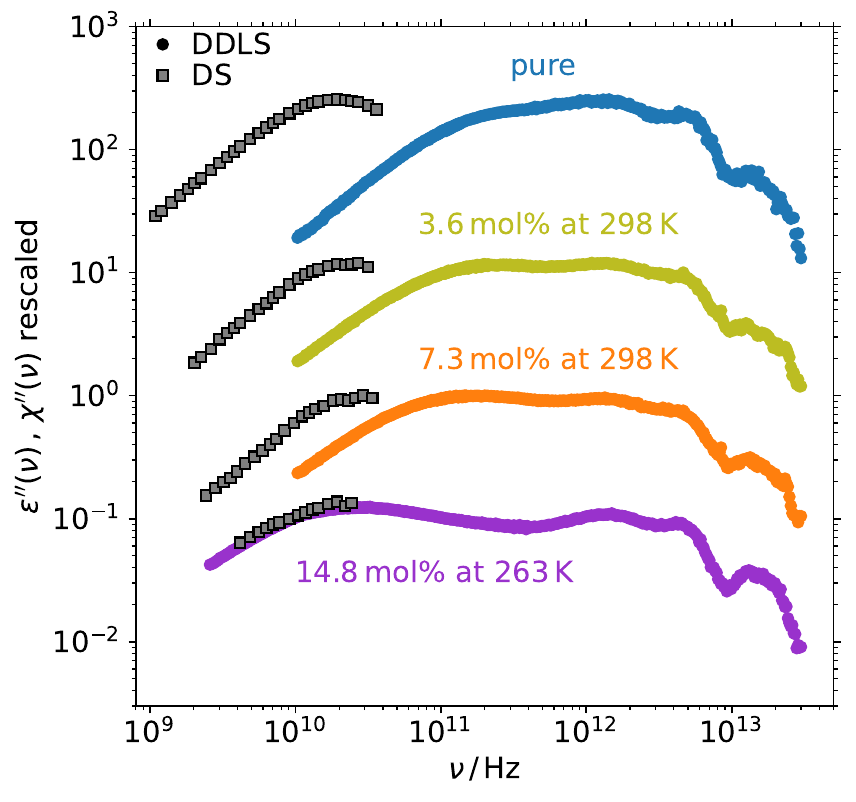}
  \caption{DDLS susceptibilities and dielectric loss spectra of pure water and some chosen aqueous lithium chloride solutions. The spectra are shifted in amplitude for better comparison.}
  \label{fig:figure3}
\end{figure}
Fig.\,\ref{fig:figure3} shows DDLS susceptibilities of pure water and some of the aqueous lithium chloride solutions in comparison to dielectric loss spectra at \SI{298}{\kelvin} and \SI{263}{\kelvin}. At low salt concentrations, the dielectric loss peaks have lower peak frequencies than the relaxation peaks of the DDLS spectra. As mentioned before, the relaxation peak in the DDLS spectra shifts to lower frequencies with increasing salt concentration. The dielectric loss peak instead shifts to slightly higher frequencies upon addition of salt, which was already observed in the literature \cite{lunkenheimer2017electromagnetic}. Therefore, the discrepancy between the peak positions obtained by the two techniques decreases with increasing salt concentration. At the concentration of \SI{14.8}{\mol\percent}, the peak positions of the spectra obtained by the two methods match rather well. This result strongly supports the hypothesis detailed in the introduction, that the BDS spectrum of neat water is dominated by dipolar cross-correlations, which are reduced by adding salt. As a consequence, the spectra of BDS and DDLS become identical at high salt concentrations. At this point, it should be noted that this result does not necessarily imply that the discrepancy between the BDS and DDLS spectra of pure water arises purely from dipolar cross correlations. While the latter will play a major role, all effects in cross- and self-correlations that lead to discrepancies between both types of spectra diminish with increasing salt concentration.\\
Since the \SI{14.8}{\mol\percent} solution is easily supercoolable, we performed additional DDLS and BDS measurements near its glass transition temperature of $\sim\SI{138}{\kelvin}$ \cite{prevel1995structural,lunkenheimer2023mysteries}. Fig.~\ref{fig:figure4} a) shows the dielectric loss as well as the DDLS spectrum at \SI{148}{\kelvin}, normalized with respect to their peak amplitude. From this comparison it becomes clear that not only the peak positions match perfectly but also the shape of the peaks is identical, at least in the deeply supercooled regime. The perfect match of BDS and DDLS spectra at \SI{14.8}{\mol\percent} suggests that dipolar cross correlation contributions are fully suppressed at this salt concentration and both methods probe the same dynamics. The inset of Fig.~\ref{fig:figure4} a) shows peak relaxation times obtained from DDLS and dielectric loss spectra of this work and from the literature \cite{lunkenheimer2023mysteries}. The solid black curve represents a fit of the Vogel-Fulcher-Tamman (VFT) equation 
\begin{equation}
\tau(T)=\tau_0\cdot e^{\frac{B}{T-T_0}},
\label{eq:VFT}
\end{equation}
where B is a constant, $T_0$ the Vogel temperature and $\tau_0$ a prefactor. The VFT fit describes the data rather well over a broad range of relaxation times, ranging from $\simeq\SI{1}{\pico\second}$ to $\simeq\SI{1}{\second}$, so about 12 orders of magnitude. This shows that the correlation times from DDLS and BDS are very similar over a much broader temperature range than discussed above. At this point it should be noted that the absence of a slow cross-correlation contribution to the dielectric loss spectra of the \SI{14.8}{\mol\percent} solution does not imply that there are no orientational cross-correlations between water-dipoles in the system. However, it suggests that the Kirkwood correlation factor $g_{\mathrm{k}}$ is close to 1, which can also result from a coexistence of parallel and antiparallel alignment of dipoles. Of course, the existence of the H-bond vibration modes in the THz regime shows that water molecules still form H-bonds, even at large salt concentrations.\\
\begin{figure}
  \centering
  \includegraphics[width=0.5\textwidth]{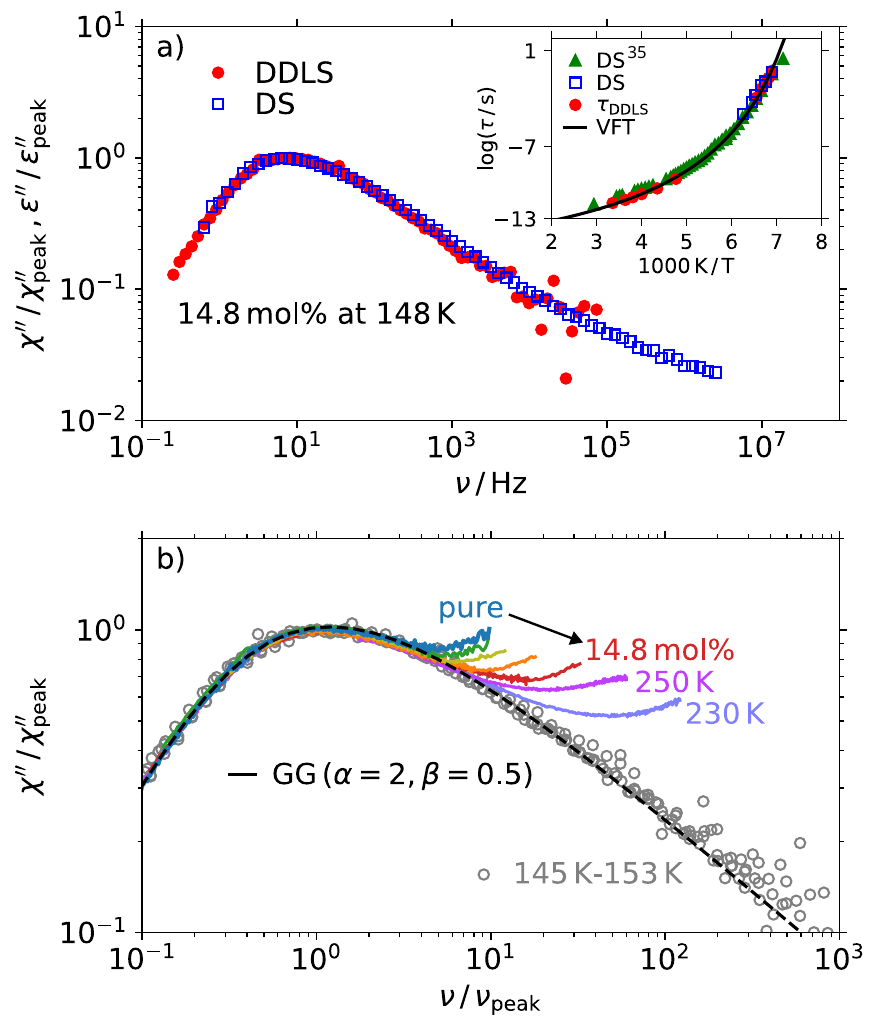}
  \caption{a) DDLS spectrum and dielectric loss of the \SI{14.8}{\mol\percent} solution at \SI{148}{\kelvin}. The inset shows an Arrhenius plot of the DDLS peak relaxation times and the BDS peak relaxation times obtained in this work (blue empty squares) and by Lunkenheimer et al.\cite{lunkenheimer2023mysteries} (green filled triangles) at the same lithium chloride concentration. The black curve represents a fit by the VFT equation. b) DDLS spectra of pure water and aqueous lithium chloride solutions at different concentrations and temperatures, normalized to the peak frequency and amplitude of the relaxation peak. The black dashed curve represents a generalized gamma distribution of relaxation times with width parameter of $\alpha=2$ and a high frequency power law parameter of $\beta=0.5$.}
  \label{fig:figure4}
\end{figure}
 Fig.~\ref{fig:figure4} b) shows the DDLS susceptibility spectra of pure water and the different lithium chloride concentrations at \SI{263}{\kelvin}, as well as the spectra obtained for the \SI{14.8}{\mol\percent} solution at different temperatures down to $\SI{145}{\kelvin}$, normalized in frequency and amplitude to the maximum of their respective relaxation peaks. The solid black curve shows the frequency domain representation of the generalized gamma distribution of relaxation times with a width parameter of $\alpha=2$ and a high frequency power law parameter of $\beta=0.5$. This distribution represents the generic shape of structural relaxation proposed in ref.\,\cite{pabst2021generic}, which is found for the DDLS spectra of very different supercooled liquids, including hydrogen bonding liquids, van der Waals liquids and even ionic liquids, as well as the dielectric loss of some (almost) non-polar substances \cite{bohmer2024dipolar, bohmer2024spectral}. The spectra superimpose rather well up to frequencies where the relaxation peak starts to superimpose with the first vibrational peak. The overlap becomes stronger with increasing temperature and decreasing concentration. However, from this it becomes obvious that the overall shape of the structural relaxation peak is independent of temperature as well as salt concentration. This independence of concentration and temperature strongly suggests that the relaxation shape is independent of the exact arrangement of water molecules in the hydrogen bond network. 

In conclusion, we confirmed, by detailed comparison of DDLS and BDS data of pure water and aqueous lithium chloride solutions in the liquid and supercooled regime, that slow dipolar cross-correlations in the dielectric response critically contribute to the discrepancy in the positions and shapes of relaxation peaks obtained by the two methods. These contributions can be reduced by the addition of salt, likely as a result of the intrusion of hydration shells of ions into the hydrogen bond network. At sufficiently large salt concentrations, the observed relaxation peaks become identical in position and shape, indicating complete suppression of dipolar cross-correlation contributions. In particular, there is no indication of a bimodal relaxation spectrum at any concentration, which would point towards dynamically distinguishable bulk-like and hydration water molecules. By contrast, the spectral shape is identical with the \emph{generic shape} previously identified in many hydrogen bonding and van der Waals liquids. This shape is observed in the presented DDLS spectra independent of temperature and salt concentration, indicating that the spectral shape is not influenced by the exact arrangement of water molecules in the system but is rather an inherent property of the structural relaxation, similar to the findings in many other systems \cite{bohmer2024spectral}. In combination, these results provide a new perspective for the analysis of relaxation data of aqueous salt solutions, which not only sheds a different light on previous analyses but will also be further tested in future studies.

\section*{Conflicts of interest}

There are no conflicts to declare.

\section*{Data availability}

The data analyzed in this study are available
from:

\section*{Acknowledgements}
Financial support by the Deutsche Forschungsgemeinschaft under grant no.\ 1192/3 is gratefully acknowledged.

\bibliography{LiCl.bib}

\end{document}